# T0 Fan-out for Back-n White Neutron Facility at CSNS

X.Y. Ji, P. Cao, T. Yu, L.K. Xie, X.R. Huang, Q. An, H.Y. Bai, J. Bao, Y.H. Chen, P.J. Cheng, Z.Q. Cui, R.R. Fan, C.Q. Feng, M.H. Gu, Z.J. Han, G.Z. He, Y.C. He, Y.F. He, H.X. Huang, W.L. Huang, X.L. Ji, H.Y. Jiang, W. Jiang, H.Y. Jing, L. Kang, B. Li, L. Li, Q. Li, X. Li, Y. Li, R. Liu, S.B. Liu, X.Y. Liu, G.Y. Luan, Y.L. Ma, C.J. Ning, J. Ren, X.C. Ruan, Z.H. Song, H. Sun, X.Y. Sun, Z.J. Sun, Z.X. Tan, J.Y. Tang, H.Q. Tang, P.C. Wang, Q. Wang, T.F. Wang, Y.F. Wang, Z.H. Wang, Z. Wang, J. Wen, Z.W. Wen, Q.B. Wu, X.G. Wu, X. Wu, Y.W. Yang, H. Yi, L. Yu, Y.J. Yu, G.H. Zhang, L.Y. Zhang, J. Zhang, Q.M. Zhang, Q.W. Zhang, X.P. Zhang, Y.T. Zhao, Q.P. Zhong, L. Zhou, Z.Y. Zhou and K.J. Zhu

*Abstract*—the main physics goal for Back-n white neutron facility at China Spallation Neutron Source (CSNS) is to measure nuclear data. The energy of neutrons is one of the most important parameters for measuring nuclear data. Method of time of flight (TOF) is used to obtain the energy of neutrons. The time when proton bunches hit the thick tungsten target is considered as the start point of TOF. T0 signal, generated from the CSNS accelerator, represents this start time. Besides, the T0 signal is also used as the gate control signal that triggers the readout electronics. Obviously, the timing precision of T0 directly affects the measurement precision of TOF and controls the running or readout electronics. In this paper, the T0 fan-out for Back-n white neutron facility at CSNS is proposed. The T0 signal travelling from the CSNS accelerator is fanned out to the two underground experiment stations respectively over long cables. To guarantee the timing precision, T0 signal is conditioned with good signal edge. Furthermore, techniques of signal pre-emphasizing and equalizing are used to improve signal quality after T0 being transmitted over long cables with about 100 m length. Experiments show that the T0 fan-out works well, the T0 signal transmitted over 100 m remains a good time resolution with a standard deviation of 25 ps. It absolutely meets the required accuracy of the measurement of TOF.

## I. INTRODUCTION

THE China Spallation Neutron Source (CSNS) is a large scientific facility that is mainly designed to carry out multidisciplinary research on material characterization making the use of neutron scattering techniques. The Back-n white neutron experimental facility for nuclear data measurement is based on the back-streaming neutrons [1][2][3][4]. It takes the advantages of a 15 ° deflection angle of the proton beam line RTBT. The neutrons produced by the spallation are canalized to the ES#1 (Experiment Station) and ES#2, which is about 56m and 76m away from the target respective, through a vacuum pipe. There will be 7 spectrometers in these two experiment stations: C6D6 detectors, a 4π-BaF2 array named GTAF-II (Gamma Total Absorption Facility II) for neutron capture cross-section (n, γ) measurements, FIXM (Fast Ionization Chamber Spectrometer for Fission Cross-section Measurement) for fission cross-section (n, f) measurements, NTOX (Neutron Total Cross-section Spectrometer) for total cross-section (n, t) measurements, LPDA (Light-charged Particle Detector Array), FINDA (Fission Neutron Spectrum Detector Array) for prompt fission neutron spectrum measurements and GAEA (Gamma spectrometer with Germanium Array) for gamma spectrum (n, n'γ/2nγ) measurements. Detailed description is given in ref. [5].

Neutron energy spectrum is a very important part for the measurement, the TOF (time of flight) is applied for the acquisition of neutron energy [5]. The kinetic energy of the neutrons is calculated relying on the measurement of the arrival

Manuscript submitted June 24, 2018. This work is supported by the National Key Research and Development Program of China (Project: 2016YFA0401602) and NSAF (No. U153011) (Corresponding author: P. Cao, e-mail: cping@ustc.edu.cn).

X.Y. Ji, P. Cao, T. Yu, L.K. Xie, X.R. Huang, Q. An, C.Q. Feng, S.B. Liu, L. Yu, Z.J. Sun, R.R. Fan and L.Y. Zhang are with the State Key Laboratory of Particle Detection and Electronics, Beijing 100049, Hefei 230026, China (hwjxy@mail.ustc.edu.cn).

P. Cao, X.Y. Ji and L.K. Xie are with the Department of Engineering and Applied Physics, University of Science and Technology of China, Hefei, China.

Q. An, C.Q. Feng, X.R. Huang, S.B. Liu, T. Yu and L. Yu are with the Department of Modern Physics, University of Science and Technology of China, Hefei, China.

H.Y. Bai, Z.Q. Cui, H.Y. Jiang, and G.H. Zhang are with the Peking University, Beijing 100871, China.

J. Bao, G.Z. He, H.X. Huang, G.Y. Luan, J. Ren, X.C. Ruan, H.Q. Tang, Qi Wang, Z.H. Wang, X.G. Wu, Q.W. Zhang, Q.P. Zhong and Z.Y. Zhou are with the China Institute of Atomic Energy, Beijing 102413, China.

M.H. Gu, X.L. Ji, Y. Li and K.J. Zhu are with the Institute of High Energy Physics, CAS, Beijing 100049, China.

Y.H. Chen, R.R. Fan, W.L. Huang, Y.C. He, H.T. Jing, W.Jiang, N. Kang, B. Li, L. Li, Q. Li, X. Li, Y.L. Ma, C.J. Ning, H. Sun, X.Y. Sun, Z.J. Sun, J.Y. Tang, Z.X. Tan, P.C. Wang, Q.B. Wu, X. Wu, Y.F. Wang, Y.J. Wu, Z. Wang, H. Yi, J. Zhang, L. Zhou and L.Y. Zhang are with the Institute of High Energy Physics, CAS, Beijing 100049, China and Dongguan Neutron Science Center, Dongguan 523803, China.

Y.F. He and P.J. Cheng are with the University of South China, Hengyang 421001, China.

Z.J. Han, R. Liu, X.Y. Liu, J. Wen, Z.W. Wen and Y.W. Yang are with the Institute of Nuclear Physics and Chemistry, CAEP, Mianyang, China.

Z.H. Song and X.P. Zhang are with the Northwest Institute of Nuclear Technology, Xi'an, China.

Q.M. Zhang and Y.T. Zhao are with the Xi'an Jiaotong University, Xi'an 710049, China.

T.F. Wang is with the Beihang University, Beijing, China.

time and the starting time of the neutrons. There is going to be a timing pulse that represents the very emission time of the neutrons, so-called T0, and the detectors will give the exact time when the neutrons arrive, and both of them are recorded down by the DAQ (Data Acquisition), actually the timing pulse is used to trigger the DAQ. The DAQ is detailed described in ref. [6].

## II. FEATURES OF T0 SIGNAL

Actually, the exact neutron emission time from the target is not possible to measure without changing it [7], but we can find other times which are relative to the neutron emission time with a fixed time interval. In Back-n white neutron source, there are three kinds of timing pulses, and we will choose the one with better time resolution depending on the final evaluation test. Another way to timing by detecting the prompt gamma burst from the target in the neutron beam line is also available at Back-n, as could be recorded by the DAQ directly, it will not discuss further. Here is the detailed description for the three kinds of T0.

The first timing pulse comes from RCS (Rapid Cycling Synchrotron) kicker magnet, supported by the control system. When the kicker magnet opens the gate to let the protons beams hit the lead target through the tube, a sensing signal will be generated by the control system, and converted to a light signal for a long distance transmission from the RCS Station to WNS CS (Control Station). In WNS Control Station the signal will be converted back to a LVTTL signal, features a range from 1μs to 10μs pulse width. It is called RCSX-T0, it is supposed to trigger the WNS readout electronics implemented in WNS CS, ES#1 (Experiment Station) and ES#2, respectively. As described in ref. [5].

The second timing pulse comes from the beam monitoring system, the sensing probe with 0.5V/A sensitivity, located beside the sixth magnet away from the 15 °bending magnet, will generate a sensing signal when the proton beams goes by. The pulse features an 86ns pulse width. The beam peak intensity reaches up to 40A when the beam power is 100kW [5]. The signal traveling along the cable to WNS Control Station, is attenuated to 8V peak amplitude when the RCS runs full power, 100kW, and 0.8V peak amplitude with the 10kW beam power. The characteristics of the signal mainly depend on the operation mode, the operation mode is detailed in ref. [5]. And it is called FCT-T0.

The last T0 signal is provided by a BaF2 detector at Collimator#1 which is used for online detection of the prompt gamma. We named it Gamma-T0.

We propose a kind solution of T0 Fan-out taking consideration of the Back-n layout. The T0 fan-out structure is indicated in Fig. 1. Both RCSX-T0 and FCT-T0 will be fan-out from WNS Control Station to ES#1 and ES#2, and they share the same cable. As for the Gamma-T0, we process the signal at Collimator#1 and drive the signal back to WNS Control Station, then the Gamma-T0 could be fan-out like FCT-T0.

## III. IMPLEMENTATION OF THE T0 FAN-OUT BOARD

The key is how to ensure the resolution of the timing pulse after such a long distance transmission for an electronic signal. There are two main factors which will increase the timing non-determinacy. First, the rise time of the signal have a significant impact on the timing resolution, especially when the cutting-edge-timing is implemented. Because the white noise always exists, the stochastic noise added to the rise edge will make the point which crosses the threshold much more nondeterministic, it is so-called time-walk effect, the slower the rise edge is, the worse the situation will be. Second, the processing model may cause extraneous jitter, it will also have bad influence on the timing resolution.

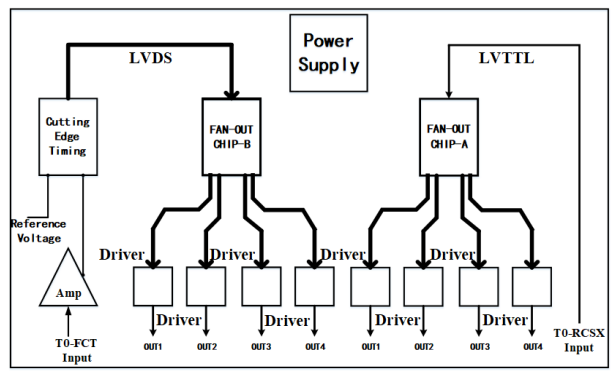

Fig. 2. Block diagram of T0 Fan-out Board

The block diagram of T0-Fanout-Board is illustrated in Fig. 2. The T0-Fanout-Board includes FCT-T0 process function block and RCSX-T0 function block. As for the FCT-T0, the signal goes through the amplifier to adapt to the comparator, cooperating with the LMH7220 to implement the cutting-edge-timing. The T0 signal has been transferred into digital signal by LMH7220, with LVDS standard. Then the LVDS remaining the information of the T0 will be fan-out by SN65LVDS104, which offers four low-noise coupling LVDS output. The next, each LVDS output signal will be driven by DS15BA101 to ensure a long distance transportation. The DS15NA101 features a very high speed signal buffer for cable driving. The RCSX-T0 is fan-out by SN65LVDS105 directly, because the chip supports a LVTTL input model and offers the same output features with SN65LVDS104. The photograph of T0 Fan-out board is shown in Fig. 3.

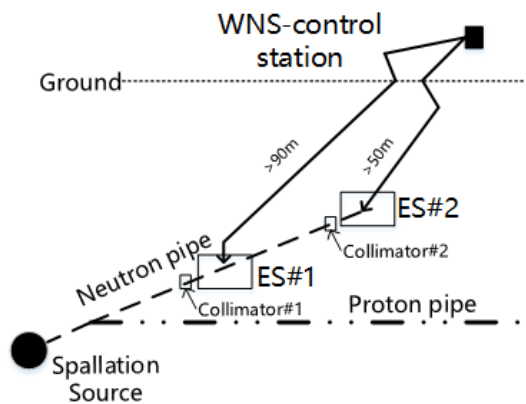

Fig. 1. Structure of ES#1, ES#2 and WNS (White Neutron Source) Control Station. The WNS Control Station is above the ground, and the coaxial cable for T0 fan-out have been laid. The longest cable is more than 100m.

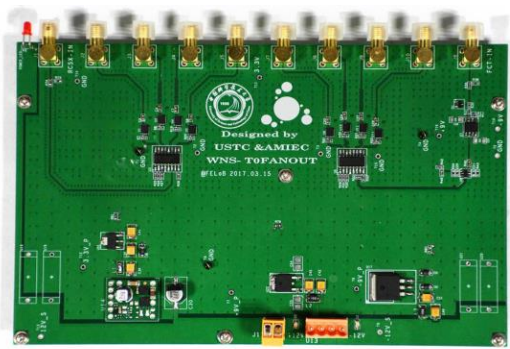

Fig. 3. The photograph of T0 Fan-out Board

## IV. TEST RESULT

The T0 Fan-out Board has been tested, focused on the difference about signal shaking between the input and output, the RMS of transport delay is an index to evaluate the performance of the board. The RCSX-T0 fan-out function block and FCT-T0 fan-out function block have been tested, respectively. The diagram of test is shown in Fig. 4.

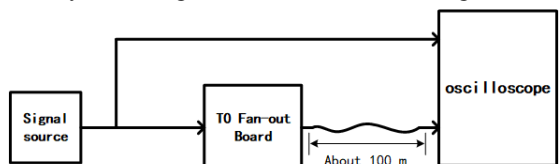

Fig. 4. The diagram of test

The RCSX-T0 test is using a signal generator to produce a cycle square wave with a leading and trailing edge of 10ns. Take the advantage of Tektronix DPO5104, we compare the input signal and the output signal which travels along a 100 meters LMR-240 coaxial-cable. The result is shown in Fig. 5, the standard deviation of the transport delay is 17.3ps, and it is enough to meet the requirement of the TOF resolution 1ns.

The FCT-T0 test is almost the same as the RCSX-T0 test except of the amplitude of the input signal, which is set to 800mV. The standard deviation of the transport delay is 24.9ps, this result is also meet the requirement.

On the other hand, another test also proves that the fan-out model meets the requirement. We use the signal generator to provide a periodic pulse to the T0 fan-out board, and calculate the jitter of the output signal which travels along a 100 meters

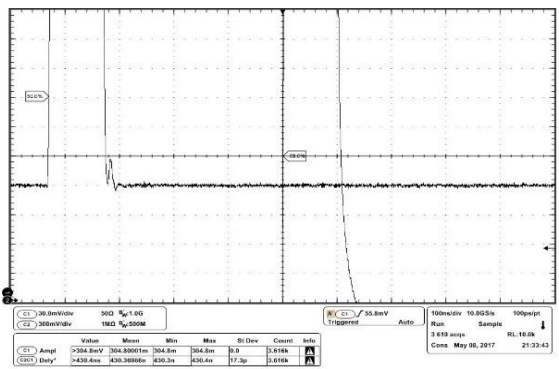

Fig. 5. RMS of transport delay test result Standard deviation 17.3ps

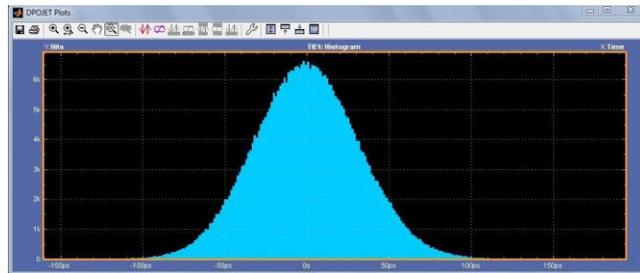

Fig. 6. The test result of the RCSX-T0 which transmits over 100m, the Standard deviation is 31.6ps,

LMR-240 coaxial-cable by using the Tektronix DPO5104. The result is illustrated in Fig. 6. The result of RCSX-T0 test is 31.6ps, indicating that the fan-out model meets the requirement.

The test of FCT-T0 jitter is also carried out, with the standard deviation of jitter 33.9ps. the T0 fan-out board is enough to satisfy the demand of the TOF measurement supported by all the test results.

## V. CONCLUSION

In this paper, we propose the idea of T0-Fanout to satisfy the needs of the TOF measurement at the Back-n white neutron source in CSNS, aiming at the problem of the T0 signal's fan-out and long distance transportation, we design the T0-Fanout-Board. The board could fan out four channel signal for each kind of T0 signal, drive the signal transmitting a long distance and ensure an extremely small jitter of the rise edge. The model has been tested to proving that it works well during the TOF experiment.